\documentclass[openacc]{rstransa}
\usepackage{subfigure}

\begin{document}

\title{Real-Time Cortical Simulation on Neuromorphic Hardware}

\author{
Oliver Rhodes$^{1}$, Luca Peres$^{1}$, Andrew G. D. Rowley$^{1}$, Andrew Gait$^{1}$, Luis A. Plana$^{1}$, Christian Brenninkmeijer$^{1}$, and Steve B. Furber$^{1}$}

\address{$^{1}$Department of Computer Science, University of Manchester, Manchester, UK}

\subject{Neuromorphic Computing, Computational Neurosicence, Spiking Neural Networks, Massively-Parallel Computing, Event-Driven Processing}


\corres{Oliver Rhodes\\
\email{oliver.rhodes@manchester.ac.uk}}

\begin{abstract}
Real-time simulation of a large-scale biologically representative spiking neural network is presented, through the use of a heterogeneous parallelisation scheme and SpiNNaker neuromorphic hardware. A published cortical microcircuit model is used as a benchmark test case, representing $\approx 1 \, \text{mm}^2$ of early sensory cortex, containing $77\text{k}$ neurons and $0.3$ billion synapses. This is the first true real-time simulation of this model, with $10 \, \text{s}$ of biological simulation time executed in $10 \, \text{s}$ wall-clock time. This surpasses best published efforts on HPC neural simulators ($3\times$ slowdown) and GPUs running optimised SNN libraries ($2\times$ slowdown). Furthermore, the presented approach indicates that real-time processing can be maintained with increasing SNN size, breaking the communication barrier incurred by traditional computing machinery. Model results are compared to an established HPC simulator baseline to verify simulation correctness, comparing well across a range of statistical measures. Energy to solution, and energy per synaptic event are also reported, demonstrating that the relatively low-tech SpiNNaker processors achieve a $10\times$ reduction in energy relative to modern HPC systems, and comparable energy consumption to modern GPUs. Finally, system robustness is demonstrated through multiple $12 \, \text{h}$ simulations of the cortical microcircuit, each simulating $12 \, \text{h}$ of biological time, and demonstrating the potential of neuromorphic hardware as a neuroscience research tool for studying complex spiking neural networks over extended time periods.
\end{abstract}

\begin{fmtext}
\end{fmtext}

\maketitle

\section{Introduction}
Neural networks provide brains with the ability to perform cognitive tasks, motor control, and learn and memorise information. These circuits are robust, fault tolerant and extremely efficient, with the human cerebral cortex consuming approximately $12 \, \text{W}$ \cite{Herculano_Houzel2011}. However, understanding these systems is a complex task requiring consideration of cellular and circuit level behaviours. While experimental measurements are readily taken at the cellular scale, gathering data from large-scale circuits is more challenging. This has led to the use of computational models to simulate the response of circuit-scale spiking neural networks (SNNs) representing brain activity. In addition to the use of SNNs as a neuroscience research tool, there is growing interest in harnessing their brain-like information processing to further the recent rapid progress in the field of deep learning and artificial neural networks. In addition to providing advanced computational capacity \cite{Maass2004}, the use of spike-based communication offers the potential to reduce significantly energy consumption, opening up potential for deployment of advanced inference/learning capabilities in edge computing devices \cite{Shi2016}, and in un-tethered/low-power neurorobotics applications \cite{Bing2018}. However, execution of SNN models is a complex process on traditional computing machinery, particularly at scale, as the long range connectivity and sparse temporal signals make traditional communication mechanisms inefficient. Communication costs therefore dominate performance, and scale nonlinearly with neural network size, slowing down simulations and increasing energy consumption of the underlying simulator.

This work presents a solution to this problem through the use of the digital neuromorphic platform SpiNNaker\cite{Furber2013}, together with software developments optimising use of the massively parallel architecture. The target use case is a benchmark cortical microcircuit model from the literature \cite{Potjans2014}, containing biologically representative numbers of neurons, connection topology, and spiking activity. This paper builds on previous simulations of the model \cite{Albada2018, Knight2018}, providing an in-depth analysis of the required memory, processing and communication within the SNN, leading to a SpiNNaker implementation capable of simulating the model at biological real time: $10 \, \text{s}$ of simulation is executed in $10 \, \text{s}$ of wall-clock time -- a feat not currently achievable using conventional HPC or GPU hardware. This opens up the potential to extend simulation times and computational neuroscience research to explore long-term effects in the brain, demonstrated here by successfully simulating in real time the target cortical microcircuit model continuously for 12 hours. In addition to real-time processing, energy performance is improved by an order of magnitude relative to previous work \cite{Albada2018}, highlighting the benefits of neuromorphic computing in the quest for large-scale neural network simulation.

To demonstrate these achievements the paper is structured accordingly: following this introduction a \emph{Background \& Motivation} section highlights the main features of the target cortical circuitry, and explores the computational requirements of simulating such a model. A \emph{Methods} section then explores how these requirements are mapped to SpiNNaker, and is followed by a \emph{Results} section detailing simulation correctness and performance. Finally, a \emph{Conclusion} section comments on findings, and makes suggestions for further improvements to the modelling process and future SNN simulator designs. 


\section{Background \& Motivation} \label{sec:background_and_motivation}
\subsection{Use Case: Cortical Microcircuit Model} \label{sec:use_case}
\begin{figure}[t!]
\includegraphics[width=\textwidth]{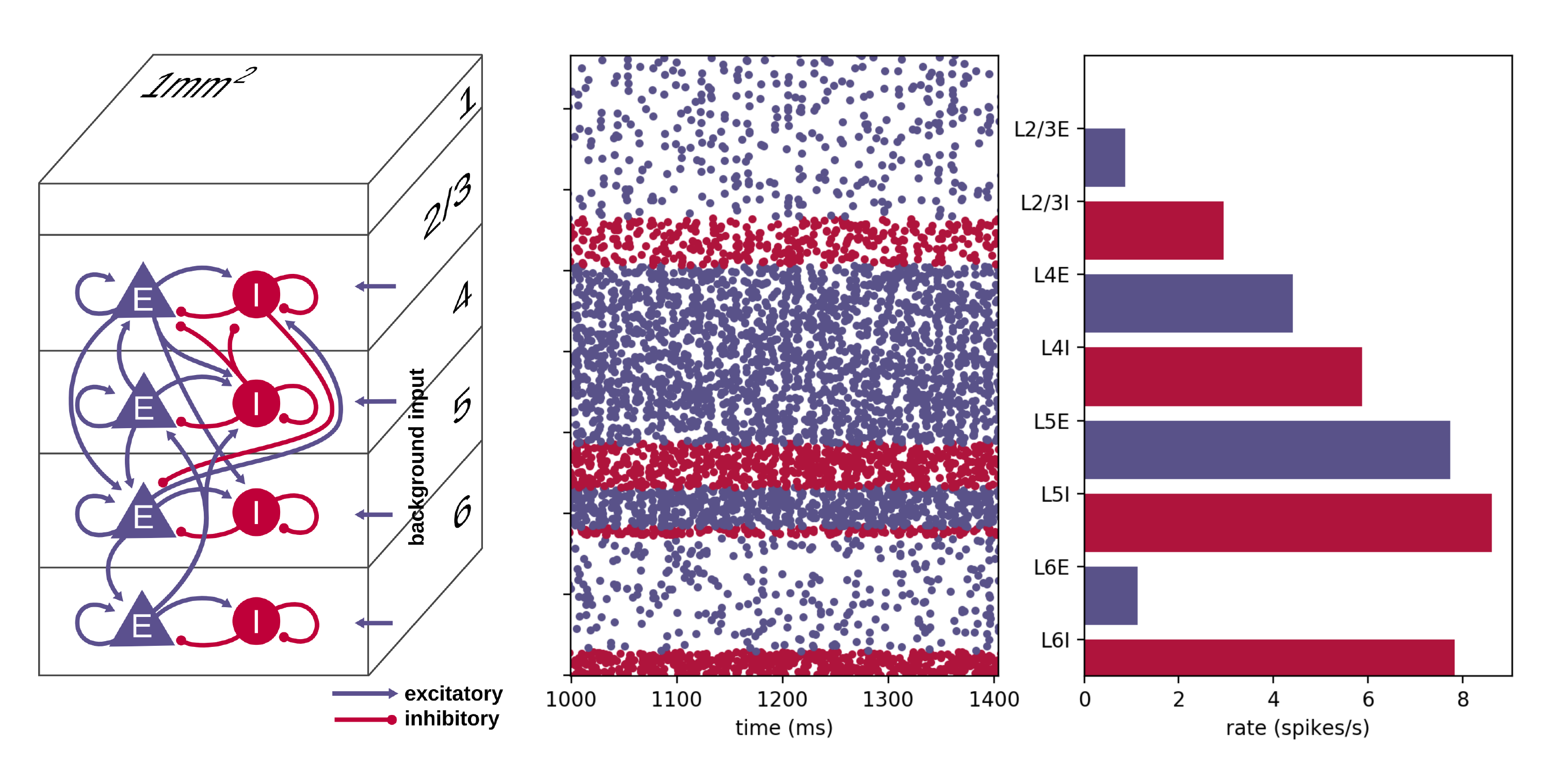}
\caption{Cortical microcircuit model: left, layered structure of excitatory and inhibitory neuron populations; middle, $0.4\,\text{s}$ of steady-state output spikes ($5\%$ of total spikes plotted for clarity); and right, layerwise mean population firing rates.}
\label{fig:microcircuit}
\end{figure}

The focus of this work is the cortical microcircuit model developed by Potjans and Diesmann \cite{Potjans2014}.  The model represents $\approx 1\,\text{mm}^2$ of generic early sensory cortex, and contains 77169 neurons and $\approx 3 \times 10^8$ synapses (see original publication \cite{Potjans2014} for a full description). Populations of neurons are arranged in layers, with each layer containing both excitatory and inhibitory neurons, with recurrent and inter-population connections as detailed in Fig.~\ref{fig:microcircuit}~left. Synaptic weights are normally distributed, with mean $\pm$ standard deviation of $351.2 \pm 35.32 \,\text{pA}$ for inhibitory source neurons, and $87.8 \pm 8.78 \, \text{pA}$ for excitatory source neurons (except connections from layer 4 to 2/3 excitatory neurons which have weights $175.6 \pm 8.78 \, \text{pA}$). Transmission delays are  normally distributed and truncated to the nearest simulation timestep $\Delta t$, with mean $\pm$ standard deviation of $1.5 \pm 0.75 \, \text{ms}$ for excitatory sources, and $0.75 \pm 0.375 \,\text{ms}$ for inhibitory neurons. Each population also receives background stimulation representing input from other brain regions. This input can either be in the form of direct current, injected directly into the neuron; or via Poisson input sources delivering synaptic input with population-dependent rates. Neurons are simulated with a single-compartment leaky integrate-and-fire unit, with current-based exponentially decaying synapses, as described by Eq.~\ref{eqn:lif}. 
\begin{align}\label{eqn:lif}
\begin{split}
\frac{\mathrm{d}V}{\mathrm{d}t} &= \frac{V-E+(I_{syn} + I_{DC})R}{\tau_m}  \qquad \mathrm{if} \,\, V > V_{\theta}, \,\, V=V_{\text{reset}} \\ 
\frac{\mathrm{d}I_{syn}}{\mathrm{d}t} &= \frac{I_{syn}}{\tau_{syn}} + \delta(t-t^j) \\ 
\end{split}
\end{align}
Membrane potential $V$, evolves with time constant $\tau_m$, relative to a resting potential $E$. Synaptic input current $I_{syn}$, is increased on spike reception, and decays with time constant $\tau_{syn}$. This synaptic current is combined with direct input current $I_{DC}$, and incorporated into the neuron membrane potential via membrane resistance $R$. When $V$ exceeds the threshold potential $V_{\theta}$, the neuron emits a spike, and its membrane potential is set to the reset potential $V_{reset}$. Simulations are performed for a total of $10\, \text{s}$ with a simulation timestep of $\Delta t = 0.1 \, \text{ms}$ for accuracy of produced spike times. Simulation output is split into an initial transient followed by steady-state activity, with the first $1 \, \text{s}$ of results discarded when postprocessing to negate the effect of initial transient activity. 
Model performance is assessed from output spiketrains, and measured in terms of firing rates across each neuronal population, the coefficient of variation of inter-spike interval, and the within-population activity correlation coefficient. These measures provide quantitative analysis of a simulation, and allow direct comparison of model results obtained under different initial conditions and from execution on a range of simulators. 

The cortical mircocircuit model is a good test case for neuromorphic hardware, as it is an accepted standard across the field of computational neuroscience, and allows direct comparison between different software/hardware simulators. It demonstrates typical firing rates and peak synaptic fan-in/-out experienced in the brain, placing constraints on processing speed, communication and available synaptic memory. The model is also defined and built in a generic high-level way typically adopted by neuroscience researchers, here using the PyNN modelling language \cite{Davison2014} (as opposed to bespoke simulator-specific coding). Synaptic connections are defined based on probability, representing a significant challenge as simplifying assumptions based on distance-dependence and proximity are unavailable to improve performance during routing and synaptic matrix generation/handling (routing must be all-to-all, encompassing all possibilities from short- to long-range connections). Successfully simulating this model is therefore a difficult challenge for neuromorphic hardware, but lays the foundations for scaling up and modelling additional and more complex brain regions in the future.

\subsection{SpiNNaker}
\subsubsection{Hardware}
SpiNNaker is a massively parallel digital neuromorphic platform for simulating SNNs. It comprises two key hardware components: the SpiNNaker chip, and within that the SpiNNaker router. A SpiNNaker chip contains 18 cores, each comprising an ARM968 processor, direct memory access (DMA) controller, network interface and communications controllers, timer, and other peripherals \cite{Painkras2013}. Cores have fast local Instruction and Data Tightly Coupled Memory (ITCM and DTCM), of $32\, \text{kB}$ and $64\,\text{kB}$ respectively. Additional chip-level storage of $128\,\text{MB}$ is provided in the form of SDRAM, with block transfer access available to all cores via their DMA controller. SpiNNaker chips are assembled onto boards, with inter-chip communication via small data packets (there is no shared memory between chips). The platform benefits from the SpiNNaker routing system: a dedicated message-passing fabric optimised for multicast transmission of many small data packets \cite{Navaridas2013}. Each chip houses a hardware router containing a list of routing table entries, each of which describe what to do with a particular packet: transmit it to a neighbouring chip(s) on one (or more) of the 6 links (N, NE, E, S, SW, W), and/or deliver it to one (or more) of the cores on the current chip (cores 1-18). The routing system therefore enables efficient multicast communication from a single source to multiple destinations -- much like action potential transmission and fan-out by a neuronal axon in biology. 

\subsubsection{Software}
The sPyNNaker software package\cite{Rhodes2018} provides a neural modelling framework for execution of SNNs on the SpiNNaker multicore hardware. SNN models are defined via the PyNN language \cite{Davison2014}, and interpreted via the Python-based SpiNNTools software package \cite{Rowley2019}. This host-based preprocessing performs model partitioning, mapping, place and route, and data generation. Model data defining neuron and synapse parameters and initial conditions, together with matrices defining synaptic connections are then loaded to the SpiNNaker machine, together with application binaries containing runtime code for execution on SpiNNaker cores. 

These applications are programmed in C, and compiled into ARM instruction code. Each application is compiled against: the SpiNNaker Application Runtime Kernel (SARK), providing access to low-level software and hardware components; and SpiN1API, a bespoke light-weight event-driven operating system. This allows linking of software callbacks to a range of hardware events, each with different priority, hence enabling construction of applications performing event-driven operations in real time. For example, during neural simulation periodic timer events are used to trigger callbacks performing regular state updates evolving neural dynamics, while the reception of spike packets is handled through high-priority packet-received events preempting all processing to ensure packet traffic is received as quickly as possible to reduce pressure on the network. Use of this operating system for neural simulation will be discussed in Sec.~\ref{sec:methods}\ref{sec:neural_simulation}, however readers are referred to previous work \cite{Rhodes2018} for detailed discussion and performance profiling of this event-based operating system in the context of neural simulation.

\subsection{Model Assessment and Previous Work}
In a recent study by Van Albada et al \cite{Albada2018}, the cortical microcircuit model was ported to SpiNNaker, and performance compared to the HPC-based NEST simulator\cite{Gewaltig2007}. The SpiNNaker implementation distributed neurons 80 per core (80 individual sources per core for Poisson inputs), as this was the minimum number enabling generation of routing tables satisfying hardware constraints. Machine allocation algorithms \cite{Rowley2019} calculated a total of 217 chips and 1924 cores were required for execution, resulting in allocation of a SpiNNaker machine consisting of 6 48-chip boards, providing a potential 288 chips and 5174 cores, with unused boards turned off during simulation. To ensure neuron and spike processing was always completed within a given timer period, the interval between timer events was set to 400000 clock cycles, yielding an effective $20\times$ slow-down relative to real time. With this configuration the model was successfully executed with no system warnings or dropped packets, and results were in excellent agreement (both at the single neuron and network level) with the HPC simulator, which was able to execute the model at $3\times$ slow-down relative to real time. Energy per synaptic event for NEST and SpiNNaker was reported respectively at $5.8 \, \mu \text{J}$ and $5.9 \, \mu\text{J}$, which is significantly higher than values reported previously for SpiNNaker: $20 \, \text{nJ}$ \cite{Stromatias2013}; and  $110\,\text{nJ}$ \cite{Sharp2012}. This reduced efficiency was explained by the simulation being sparsely distributed over the 6-board SpiNNaker machine, causing baseline power to be amortised across many fewer synaptic events\cite{Albada2018}; and due to the simulation being slowed down by a factor of $20$, causing the system to be powered for a longer period of time. A separate study explored execution of this model on GPU hardware \cite{Knight2018}, using bespoke libraries optimised for SNN simulation \cite{Yavuz2016}. Performance was evaluated on a range of GPUs with best performance from a Tesla V100 achieving processing speed of $2\times$ slow-down relative to real time, and an estimated $0.47 \, \mu \text{J per synaptic event}$. It is noted that a Jetson TX2 GPU was able to achieve superior energy consumption of $0.3 \, \mu \text{J per synaptic event}$, however simulation processing speed dropped to $25\times$ slow-down relative to real time. It is also noted that performance was evaluated for simulation models which fit entirely within GPU memory, and hence these figures are unlikely to hold if simulations required multiple GPUs to be connected and data shared between them during excecution. 

While the cortical microcircuit model was executed successfully on SpiNNaker, performance of the neuromorphic hardware for this biological model was lower than expected. While this can be attributed in part to differences between the model and initial SpiNNaker design targets \cite{Furber2013}, certain parts of the simulation process were sub-optimal. This work therefore looks to understand how a model such as this can be better fit to the SpiNNaker hardware, and the lessons which can be learned in order to further the design of future neuromorphic systems. A starting point for this research is therefore to quantify model requirements from the perspective of a simulation platform. Results from simulation of the cortical microcircuit model on the HPC-based NEST simulator (running in precise mode for maximum accuracy) are designated as the benchmark throughout, enabling performance comparison in terms of real-time execution and energy efficiency. 

\begin{figure}[t!]
    \centering
    \includegraphics[width=\textwidth]{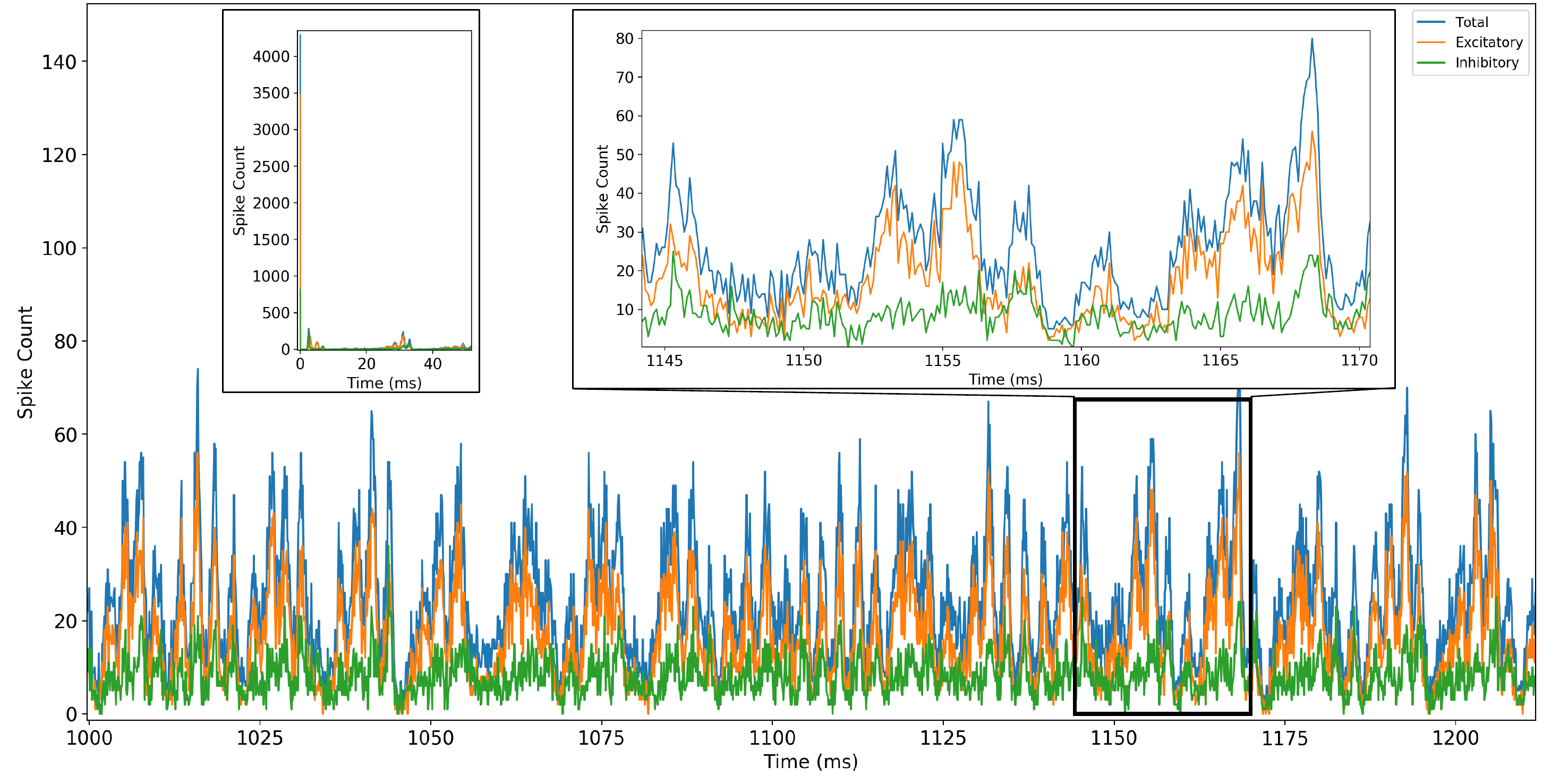}
    \caption{Analysis of cortical microcircuit output activity simulated with NEST: total, excitatory and inhibitory, spikes produced per simulation timestep. Left inset shows initial transient response, while right inset details steady-state oscillations.}
    \label{fig:model_assessment}
\end{figure}

A fundamental requirement of a \emph{true} real-time simulator is dealing with all spikes produced by the simulated network within a simulation timestep. This is different from \emph{mean} real-time processing: where an entire simulation (containing many timesteps) is executed within the same total wall-clock time. This is an important consideration when building an SNN simulator for interaction with external real-time systems, as unexpected behaviour may occur and information may be lost, due to local variations in processing speeds causing phase shifts between the two systems. Such activity is demonstrated clearly by the cortical microcircuit model, when comparing mean firing rates to the variable instantaneous rates produced by network oscillations. For example, while Fig.~\ref{fig:microcircuit}~(right) shows the mean layer-wise firing rates for the model, the total spikes produced for each simulation timestep are plotted in Fig.~\ref{fig:model_assessment} (both plots produced from baseline NEST simulation results).
This demonstrates significant variations in counts of model spikes per timestep from the initial transient phase of the simulation through to steady-state. Therefore, designing a system to cope with the mean rate may be sufficient to handle mean spike rates at \emph{mean} real time, however it will slow down or lose information during oscillations peaks of higher activity. Connectivity in the cortical microcircuit model obeys Dale's law \cite{Strata1999}, meaning all excitatory neurons make only excitatory connections, and inhibitory neurons only inhibitory connections. It is therefore interesting to split the total number of spikes produced per timestep into excitatory and inhibitory counts. The inset shows results from $1145 < t < 1170 \, \text{ms} $, and demonstrates that population activity oscillations are typically in phase, but of different magnitude depending on neuron type. Mean activity in this period produces $17.0$ excitatory and $8.3$ inhibitory spikes per timetstep, with peaks of $55$ and $38$ spikes respectively. The initial transient response of the model is extreme, with initial conditions causing production of over $4000$ spikes in the first timestep. The SNN quickly damps this activity leading to steady state behaviour after two subsequent regions of high spike output (Fig.~\ref{fig:model_assessment}, left inset).

Two additional features compromising SpiNNaker performance in previous work \cite{Albada2018} were the mechanisms handling network input and synaptic delays. Network input can be provided to the model in two ways: DC current injected into a neuron; or Poisson generated spike-based input delivered to neurons through their synapses. When simulating the Poisson version of the model on SpiNNaker, individual cores are loaded with applications capable of running a Poisson process, and sending packets targeting postsynaptic neurons (see Fig.~\ref{fig:mapping_and_data_transfer}~top). For the high rates of background activity present in the cortical microcircuit model, this resulted in certain cores receiving in excess of 300 spike packets per simulation timestep, placing huge strain on both the SpiNNaker communications fabric and spike processing pipeline. Additional load was also introduced through handling of synaptic delays which exceeded the length of on-core synaptic input buffers. This caused spikes with synaptic delays of over $16\Delta t$ $(1.6 \,\text{ms})$ to be delivered to \emph{delay extension} cores \cite{Rhodes2018, Albada2018}, which stored spikes for a given period before forwarding to postsynaptic cores at the appropriate time (see Fig.~\ref{fig:mapping_and_data_transfer}~top). As not all synapses required this delay, packets were routed directly between pre and postsynaptic cores, in addition to via the delay extension core, greatly increasing network traffic and reducing efficiency as spike packets targeted fewer postsynaptic neurons on a single core. Together these features significantly increased the number of spike packets arriving at individual cores, with increases of over $300\,\text{spike packets}$ relative to that shown in Fig.~\ref{fig:model_assessment}. 

The remainder of this work therefore explores how to improve performance of the SpiNNaker implementation of the cortical microcircuit model, through better mapping to SpiNNaker hardware and an alternative formulation of neural processing software.

\section{Methods} \label{sec:methods}
This section explores an alternative SpiNNaker software configuration to address the cortical microcircuit model requirements described and analysed in Sec.~\ref{sec:background_and_motivation}. A heterogeneous parallelisation approach is presented, capable of scaling with the demands of large-scale biologically representative SNN simulation. This is followed by implementation of a board-to-board timer alignment protocol, and discussion of techniques for measuring system energy consumption. 

\subsection{Parallelisation of Neural Simulation}\label{sec:neural_simulation}
This section describes a heterogeneous parallelisation of SNN operations, where ensembles of cores are used to address different neural processing tasks. It demonstrates how SpiNNaker hardware is assigned to parallelise neural processing operations, extending concepts explored by Knight et al \cite{Knight2016}, and applying them within the core SpiNNaker software stack. The programming model description includes: mapping of processing tasks to the SpiNNaker chip, together with updated communication strategies; the different memory structures and how they are shared between cores; and real-time event-driven execution.

\subsubsection{Mapping \& Communication}\label{sec:communication}
\begin{figure}[t!]
    \centering
    \includegraphics[width=\textwidth]{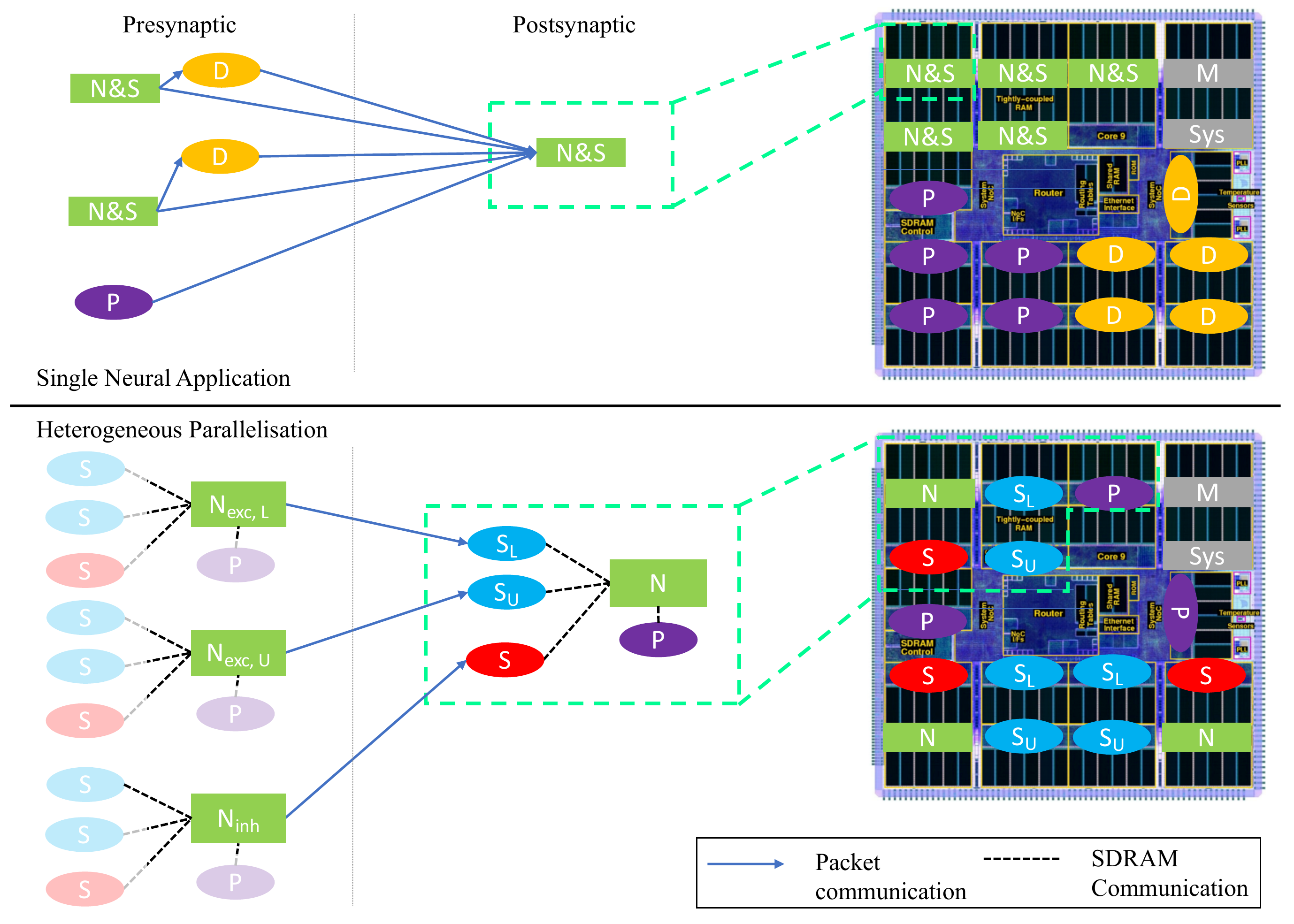}
    \caption{Mapping of pre and postsynaptic application cores to a SpiNNaker chip: top, single neural application combining neuron and synapse processing (N\&S), together with Poisson (P) and Delay Extension (D) cores, using entirely packet-based communication; bottom, neural processing ensemble containing dedicated neuron (N), synapse (S) and Poisson cores (P), with local communication via shared memory. Monitor (M) and system (Sys) cores are required for correct chip/machine operation, but are not used directly by a simulation.}
    \label{fig:mapping_and_data_transfer}
\end{figure}

A fundamental requirement of a \emph{true} real-time simulator targeting the cortical microcircuit model, is handling of the peak spike rates presented in Fig.~\ref{fig:model_assessment}. Current SpiNNaker SNN simulation software \cite{Rhodes2018} maps a neural sub-population to a single core (see Fig.~\ref{fig:mapping_and_data_transfer} top). This core is responsible for updating the state of all its neurons, and processing all incoming spike packets delivering synaptic input. 
With this approach, the core must handle the incoming packet peaks associated with the 'Total' curve in Fig.~\ref{fig:model_assessment}, in addition to spike packets from \emph{delay extension} cores (representing synaptic events with delays too large to be handled by the postsynaptic core synaptic input buffers) and spike packets from Poisson sources representing background input. While it is possible to reduce the number of neurons simulated on a core to reduce the time required to update their state, this does not reduce significantly the number of incoming spike packets, which typically dominates core processing activity. Furthermore, reducing the number of neurons per core actually reduces spike processing efficiency, as the fixed costs of turning a spike into neural input are amortised over fewer individual neuron contributions \cite{Rhodes2018}.   

This work therefore employs an alternative communication strategy, enabling parallelisation of spike packet reception, and replacement of local packet-based communication with data transfer through shared memory. This model is termed `heterogeneous parallelisation' as cores now act as part of a cooperative ensemble as shown in Fig.~\ref{fig:mapping_and_data_transfer} bottom. This ensemble is made up of: a neuron core (green) responsible for performing the neural state update; a Poisson input core (purple) providing background input; and synapse cores for processing incoming spike packets (blue for excitatory, red for inhibitory). Synaptic and Poisson input is now shared with neuron cores through shared memory, greatly reducing the incoming spike packet traffic to a single core. Furthermore, incoming spike packets now target synapse cores based on their type, with inhibitory spikes arriving at inhibitory synapse cores, and vice versa for excitatory. Inhibitory synapse cores are therefore required to handle only the reduced peak spike rates marked 'Inhibitory' in Fig.~\ref{fig:model_assessment} (as opposed to those marked `Total'), effectively parallelising synaptic processing. This concept is extended for excitatory synapse cores, as the 'Excitatory' peaks in Fig.~\ref{fig:model_assessment} are effectively double the `Inhibitory' level. This load balancing is performed by routing spikes produced from the lower half of excitatory presynaptic populations to the lower excitatory synapse core (marked $\text{S}_{\text{L}}$), and spikes from the upper half of excitatory presynaptic populations to the upper excitatory synapse core (marked $\text{S}_{\text{U}}$). As neurons within a population are equally likely to emit spikes (there is no within-layer spatial information in the cortical microcircuit model) this mechanism effectively delivers half the peak 'Excitatory' load from Fig.~\ref{fig:model_assessment}, to each excitatory synapse core, further parallelising spike processing. This distribution of incoming packets between multiple receivers is a key difference to the previous sPyNNaker software implementation \cite{Rhodes2018}, and means that all spikes are no longer routed to every receiver. It is noted that this mechanism could be further extended, through the addition of extra synapse cores to the ensemble, to cater for SNN models with higher fan-in or peak spike rates.

\begin{figure}[h!]
    \centering
    \includegraphics[width=0.45\textwidth]{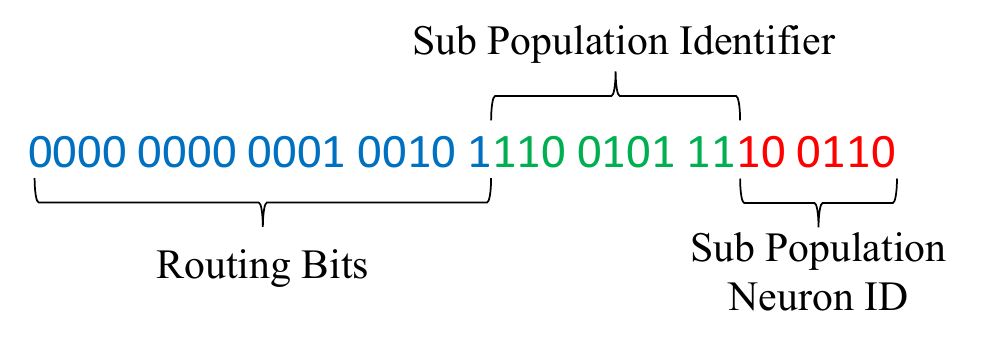}
    \caption{32-bit Multicast packet key structure (no payload) -- field sizes are representative and are adjusted based on source population size and target parallelisation.}
    \label{fig:packet_key_structure}
\end{figure}

Mapping and partitioning constraints are applied during conversion of the high-level PyNN-defined network describing the structure in Fig.~\ref{fig:microcircuit}~left, into a form suitable for loading to a SpiNNaker machine. These constraints ensure an ensemble of neural processing cores (dashed green box in Fig.~\ref{fig:mapping_and_data_transfer} bottom) are all mapped to the same chip, and can therefore share data through common SDRAM. Neuron-to-neuron communication is performed through multicast data packets, following the same approach described in \cite{Rhodes2018}. These packets are source routed based on the AER model\cite{Mead1989}, where spike packets carry only information about the source neuron, and are decoded at the receiving end to ascertain target synaptic contributions. In this work, packet key generation is updated to assist processing on the postsynaptic core, and now follows the structure defined in Fig.~\ref{fig:packet_key_structure}.
From least to most significant bits, a 32-bit key is comprised of: 6 bits to define neuron ID between 0 and 63 within the presynaptic core; 9 bits to define the location of the sub-population within the global presynaptic population (allowing up to 512 64-neuron partitions, and hence PyNN-level populations of up to 32768 neurons); and 17 routing bits which the key generator is free to use to optimise route compression and hence minimise routing table entries. Note that use of the sub-population location to improve spike processing performance is discussed further in Sec.~\ref{sec:methods}\ref{sec:neural_simulation}\ref{sec:memory_and_processing}.

\subsubsection{Memory \& Processing} \label{sec:memory_and_processing}
This section describes how an ensemble of cores cooperates to process neuron state updates and synaptic input. Shared data structures used at each stage of the process are presented, together with their locations and exchange mechanisms. SpiNNaker chip memory is allocated $64\,\text{kB}$ per core (DTCM), with an additional chip-level $128 \, \text{MB}$ of shared memory (SDRAM). Due to the highly-connected cortical microcircuit topology, it is not possible to store synaptic matrices defining network connectivity within synapse core DTCM. These data structures are therefore kept in SDRAM, and retrieved on spike arrival as discussed in previous work \cite{Rhodes2018}. 

\begin{figure}[t!]
    \centering
    \includegraphics[width=\textwidth]{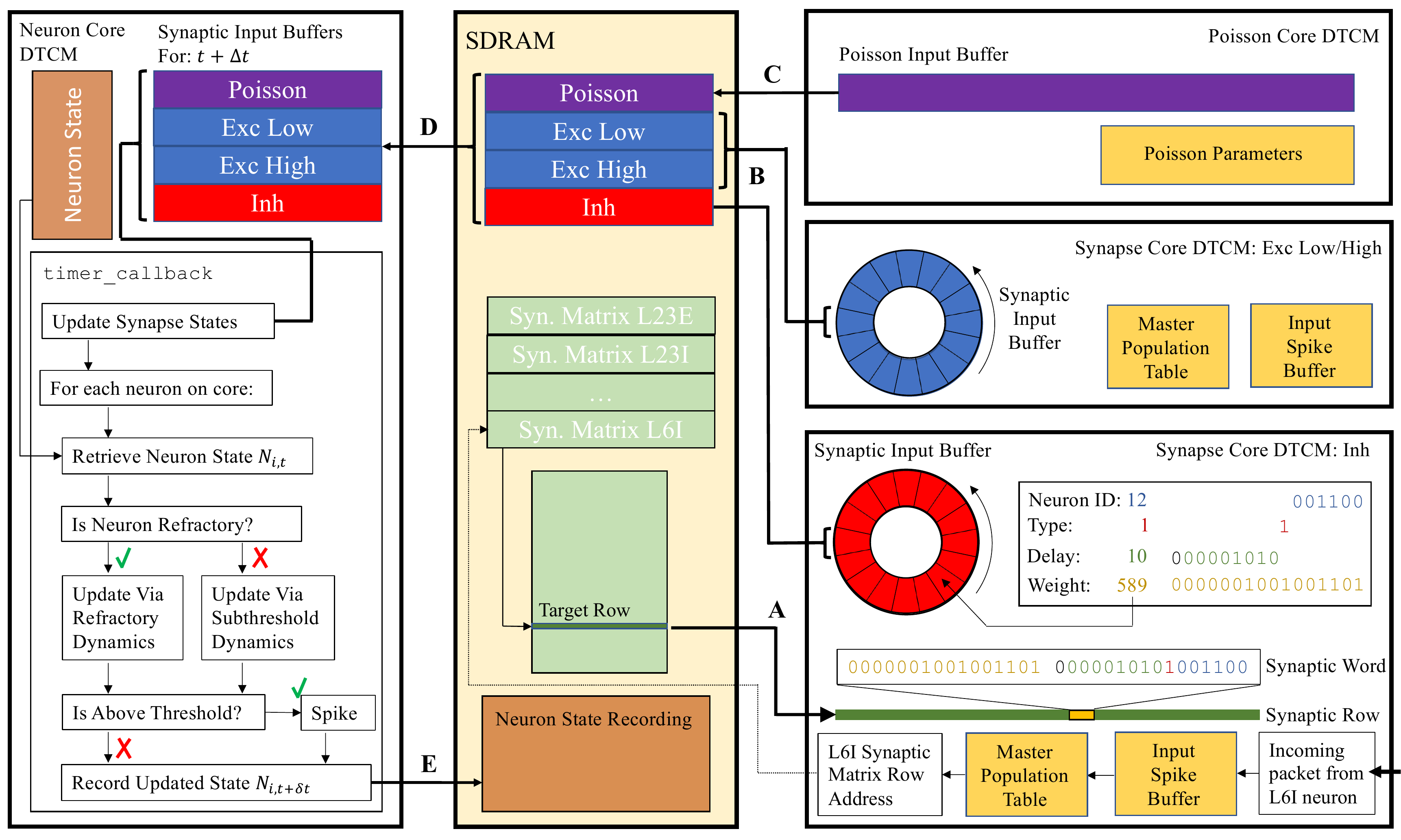}
    \caption{On-chip memory use and data structures for SNN simulation. Cores use shared memory for transfer of local information within a neural processing ensemble. Bold capital letters mark DMA transfers corresponding to the same labels in Fig.~\ref{fig:neural_processing}.}
    \label{fig:neu_syn_mem}
\end{figure}

Figure~\ref{fig:neu_syn_mem}, details memory use for an ensemble of neural processing cores, together with interactions with specific neural processing tasks. 
The lower right-hand side of Fig.~\ref{fig:neu_syn_mem}, shows an inhibitory spike packet arriving at the inhibitory synapse core, triggering the spike processing pipeline. This triggers a software callback (see Sec.~\ref{sec:methods}\ref{sec:neural_simulation}\ref{sec:realtime_processing}) which uses the packet key (see Fig.~\ref{fig:packet_key_structure}) in a binary search of the \emph{master population table} (see \cite{Rhodes2018} for details). This look-up results in the start address in SDRAM of the appropriate synaptic matrix -- here for the L6I population. The synaptic matrix is row-indexed by presynaptic neuron, with individual rows comprised of multiple synaptic words, each characterising a single connection from the presynaptic neuron to a single postsynaptic neuron. In an update from previous work \cite{Rhodes2018}, the packet-key neuron ID and sub-population identifier fields are combined to calculate a memory jump to the target row in the population-level synaptic matrix. This mechanism removes the need of an individual \emph{master population table} entry for each sub-population partition, reducing the length of the \emph{master population table} (and hence binary search times performed on the arrival of every spike packet) from $>1000$ to $4$ entries (corresponding to the four source excitatory PyNN populations on excitatory synapse cores, and the four source inhibitory PyNN populations on inhibitory synapse cores). Once a target row has been located, its contents are copied via DMA to synapse core DTCM for processing. Each synaptic word is decomposed into: target neuron ID, synapse type, synaptic delay and weight (see Fig.~\ref{fig:neu_syn_mem}). Synaptic input buffers, organised one per neuron and containing 255 slots, allow for accumulation of synaptic input for delivery over the next $255\Delta t =25.5 \,\text{ms}$. The separation of neural processing into neuron and synapse cores frees DTCM memory for synapse cores, facilitating these extended buffers, and the removal of delay extension cores used previously \cite{Albada2018}. This greatly reduces spike packet traffic, improving system efficiency and spike processing capacity. At the end of the timer period marking a simulation timestep, slots for each neuron containing input for timestep $t+\Delta t$ for all synaptic input buffers are transferred to SDRAM via DMA~\textbf{B}, and pointers updated incrementing the buffers one slot forward in time. 

Poisson background input is delivered to all neurons in the cortical microcircuit, each with model-defined rates and weights. In the approach presented here, this input is generated by a dedicated core (top right Fig.~\ref{fig:neu_syn_mem}), and transferred to the neuron via SDRAM. As this input is intended to be delivered through a synapse, undergoing accumulation and decay according to Eq.~\ref{eqn:lif}, this input is generated and copied to SDRAM (DMA~\textbf{C}) in a similar fashion to a synapse core. The Poisson core maintains a 16-bit Poisson input buffer for each associated neuron (totalling 64 per Poisson core for the implementation presented here), along with a Poisson source following a Poisson process as described in previous work \cite{Rhodes2018}. In parallel with a neuron core performing its state update, the Poisson core updates the state of each Poisson source, and determines whether it should emit a spike, and if so how many. This count is then used to add a weight value (scaled according to synaptic weights) to the Poisson input buffer for this source. On completing the update the entire buffer is copied to SDRAM, and then cleared ready for use in the next timestep. This mechanism for handling Poisson input through shared memory greatly reduces spike packet traffic, without compromising the functionality or flexibility of the system.

The left-hand side of Fig.~\ref{fig:neu_syn_mem}, shows a neuron state update, where the synaptic and Poisson inputs for this timestep for all neurons are transferred from SDRAM to neuron core DTCM (DMA~\textbf{D}), and used to update the synaptic state and membrane potential dynamics of each individual neuron (using Eq.~\ref{eqn:lif}, as discussed in \cite{Rhodes2018}). Neuron cores must therefore hold in DTCM the state variables and parameters defining the neuron and synapse models, together with recording buffers. Compared to previous approaches \cite{Rhodes2018}, the splitting of synapse and neuron cores splits memory requirements, leaving additional capacity on both cores. While this was used to model extended synaptic delays on synapse cores, it could be used to increase neuron capacity on neuron cores. However, as discussed in Sec.~\ref{sec:methods}\ref{sec:neural_simulation}\ref{sec:realtime_processing}, the system is currently processor constrained, setting capacity at $64 \, \text{neurons per core}$. On state update completion, user-requested neuron state variables are recorded to SDRAM (DMA~\textbf{E}), for subsequent extraction via the host-based software described in \cite{Rowley2019}. Note that for SpiNNaker simulations of the cortical microcircuit recorded data comprises output spike times of all model neurons. 

\subsubsection{Real-Time Event-Driven Processing} \label{sec:realtime_processing}
\begin{figure}[t!]
    \centering
    \includegraphics[width=\textwidth]{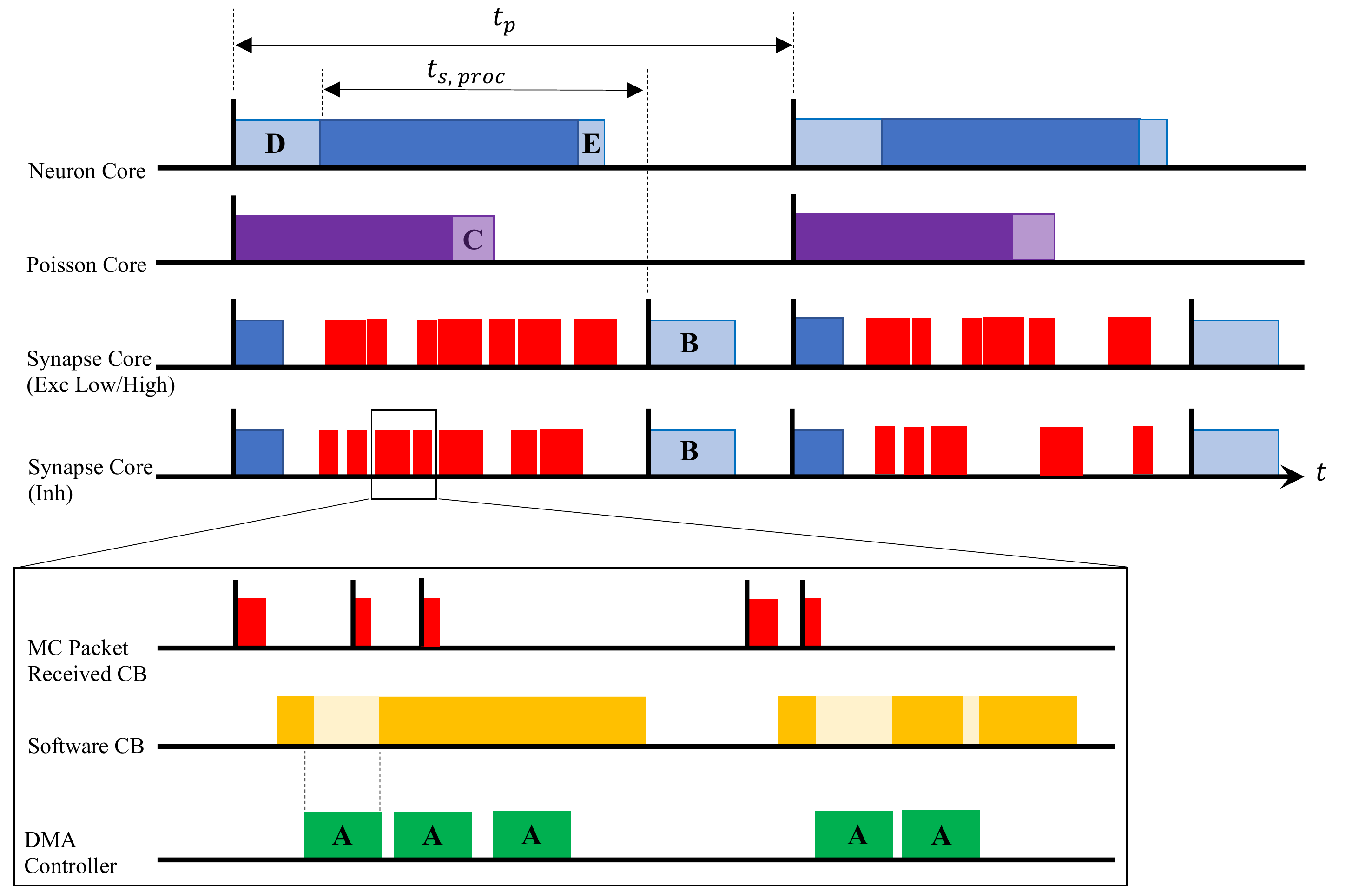}
    \caption{Event-based real-time execution of a neural processing ensemble. Bold capital letters mark DMA transfers corresponding to the same labels in Fig.~\ref{fig:neu_syn_mem}}
    \label{fig:neural_processing}
\end{figure}

This section details real-time operation of neural simulation, including how the communication through packets and shared memory described in the previous sections fits within the event-based processing framework. Figure~\ref{fig:neural_processing}, shows the ensemble of cores from Fig.~\ref{fig:mapping_and_data_transfer}, along with how their respective operations are distributed in time. Black vertical lines and associated coloured rectangles correspond respectively to hardware events and software callbacks. 
The system is clock-driven, with each core working off its own timer, but with all timers on the same SpiNNaker board effectively aligned due to sharing the same crystal source (see Sec.~\ref{sec:methods}\ref{sec:clock_alignment}). Each core responds to a periodic timer event, through an associated timer callback, as described in previous work \cite{Rhodes2018}. For real-time processing, with a simulation timestep of $\Delta t = 0.1 \, \text{ms}$, the timer period is also $t_p = 0.1 \, \text{ms}$, giving $20000\,\text{clock cycles}$ between timer events on the $200 \, \text{MHz}$ SpiNNaker ARM968. This timer provides the alignment necessary for performing DMAs \textbf{A}-\textbf{E} as demonstrated by Fig.~\ref{fig:neural_processing} and detailed below.

\paragraph{Neuron core} immediately requests DMA~\textbf{D} at the beginning of its timer callback (pale blue region in Fig.~\ref{fig:neural_processing}), transferring the Poisson and synaptic input required for the subsequent synapse and neuron state updates. These updates constitute the remainder of the software callback (dark blue bar in Fig.~\ref{fig:neural_processing}), and on completion updated state variables are recorded in SDRAM (DMA~\textbf{E}), marking the end of the callback. Above-threshold neurons emit spike packets as they are updated, with packet keys constructed using the neuron ID (see Fig.~\ref{fig:packet_key_structure}), and packets sent via the NoC to the chip router for transmission to all target synapse cores across the SpiNNaker machine. Individual neuron updates are measured at $1.05 \, \mu\text{s}$ per neuron, with the initial DMA read of $512 \, \text{bytes}$ taking a maximum of $2.68 \, \mu \text{s}$. This measurement assumes the Poisson version of the cortical microcircuit, and mapping of three ensembles (and hence three neuron cores) to a SpiNNaker chip. These measurements specify that 64 neurons can be simulated on a neuron core with $100 \, \mu \text{s}$ timer period. This ensures all neurons will have updated their state and sent any spike packets before $70 \, \mu \text{s}$ of the timer period have elapsed, ensuring all spike packets are delivered to postsynaptic synapse cores within the spike processing window $t_{s,proc}$.

\paragraph{Poisson core} uses its timer callback to update the state of all Poisson sources, and add the appropriate multiple of the associated scaled synaptic weight to the Poisson input buffer slot. Sources in the cortical microcircuit model have rates in the range $12.8 \leq f \leq 23.2 \, \text{kHz}$, meaning they are likely to produce multiple spikes per simulation timestep. At these frequencies, the average update time for 64 Poisson sources, combined with the buffer transfer time to SDRAM is $63.81 \, \mu\text{s}$. This transfer (DMA~\textbf{C}) is indicated by the pale purple region in Fig.~\ref{fig:neural_processing}, and occurs approximately mid-way through the timer period, minimising SDRAM contention with reads/writes by other cores in the ensemble. Note that for the DC version of the model, this core is omitted from the ensemble, and the buffer is not transferred into the neuron core within DMA~\textbf{D}. 

\paragraph{Synapse cores} use their timer callback to schedule a second timer event at a fixed distance in time from the end of the timer period. This event triggers a software callback responsible for transferring the latest synaptic input buffer to SDRAM (DMA~\textbf{B}). There is some core-to-core variation in SDRAM access times across the SpiNNaker chip, as multiple cores contend for SDRAM write bandwidth (potentially 9 synapse cores writing per chip -- see Fig.~\ref{fig:neu_syn_mem}), with mean (max) DMA times for writing $128 \, \text{bytes}$ measured at approximately $5\, (7.2) \, \mu \text{s}$. This second timer event is therefore scheduled $10 \, \mu \text{s}$ before the end of the timer period, ensuring DMA~\textbf{B} is complete before the associated neuron core reads this memory during its next timer callback. The window between the neuron core beginning the neuron state update (after completing DMA~\text{D}) and the synapse core writing this data (DMA~\textbf{B}), therefore determines $t_{s,proc}$, the time available for processing incoming spikes (see Fig.~\ref{fig:neural_processing}). If at the end of this window there are unprocessed entries in the input spike buffer, the second timer callback interrupts spike processing (due to elevated priority), and flushes the incoming spike buffer discarding any unprocessed spikes. This ensures real-time execution is honoured regardless of SNN activity, guarantees \emph{true} real-time simulation at the expense of lost information during periods of high activity. To monitor performance, the maximum number of spikes flushed in a single timestep is recorded throughout a simulation. 

Individual spike processing is pipelined according to methods described in previous work \cite{Rhodes2018}. Spike packet arrival triggers a high-priority packet-received event linked to a fast packet-received callback, which buffers the packet key for subsequent processing (`MC Packet Received CB' in Fig.~\ref{fig:neural_processing}). If a packet arrives while the spike processing pipeline is inactive, a software callback is scheduled to initiate spike processing (`Software CB' in Fig.~\ref{fig:neural_processing}). This instigates look-up of the spike key in the \emph{Master Population Table}, and transfer to core DTCM of the synaptic matrix target row from SDRAM (as described in Sec.~\ref{sec:methods}\ref{sec:neural_simulation}\ref{sec:memory_and_processing}). On completion of the DMA, synaptic words are extracted from the row, and converted into synaptic input buffer contributions as demonstrated in Fig.~\ref{fig:neu_syn_mem}. Once the row has been processed, the input spike buffer is checked to see if subsequent spike packets have arrived while processing the first, and the row retrieval and processing loop is repeated until the buffer is empty (continuation of solid yellow block in Fig.~\ref{fig:neural_processing} inset). At this point the pipeline becomes inactive, and the core idles until interrupted (either by subsequent packet-received or timer events). Readers are invited to review previous work \cite{Rhodes2018}, for an in-depth description and performance analysis of this processing pipeline. 

Several updates to the spike processing pipeline are introduced in this work to optimise performance and increase spike-processing capacity. First, due to reduced processing requirements of a synapse core in the heterogeneous parallelisation approach relative to a single neural application (ensemble structure shown in Fig.~\ref{fig:mapping_and_data_transfer}), it is possible to simplify callback priorities and structure. Here spike processing is treated as a background task, freeing up the need to use an interrupt-based response to DMA completion when transferring synaptic data. After requesting a synaptic row (DMA~\textbf{A} in Fig.~\ref{fig:neu_syn_mem}), the synapse core polls its DMA controller status bit to check for transfer completion. This simplification not only removes hazards, but improves performance as the interrupt service routine no longer needs to be processed -- a costly operation considering it must be processed once for every received spike. The heterogeneous programming model also removes the need for queuing of multiple DMAs by a single core, simplifying the API for interacting with the controller as no software queue needs to be maintained or managed. This reduces latency on software functions triggering the DMA, and improves performance without compromising functionality. Several additional optimisations have been made to the framework for software routing: the process whereby a packet key is converted into the target synaptic row address in SDRAM. As discussed previously, the updated packet key structure of Fig.~\ref{fig:packet_key_structure}, enables significant reductions in size of the \emph{master population table}, greatly reducing average binary search times. However, through recognition of the cortical microcircuit structure and topology, it is possible to further optimise performance. For example, while the sPyNNaker software package can handle a generic SNN model with multiple projections between populations, the cortical microcircuit model has only a single receiving projection between each population. This information is used to simplify code handling address look-up, acknowledging the fact that only a single address will be returned, and removing multiple loops and conditional statements accordingly. Overall these updates enable a single spike targeting a single neuron to be processed in $3.55\, \mu \text{s}$, a $31\, \%$ reduction relative to previously published results \cite{Rhodes2018}. It is useful to contrast this number with the spike processing window, which effectively begins when neurons start sending spike packets, and ends when the second timer event is triggered, giving it range: $3.73 \, \mu \text{s} < t_{s,proc} < 90 \, \mu \text{s}$.

\subsection{SpiNNaker Machine Clock Alignment} \label{sec:clock_alignment}
SpiNNaker machines are comprised of multiple 48-chip SpiNNaker boards, with board-to-board communication via data packets. All cores on the same board share a common crystal governing clock speed, and therefore have aligned timers. However, crystal manufacturing variability leads to small differences in clock speed between boards, and hence cores on different boards may have slightly different timer speeds. When running a timer driven simulation, these variations can lead to significant clock drift between boards, particularly when running long-duration simulations. The heterogeneous programming model of Sec.~\ref{sec:methods}\ref{sec:neural_simulation} is sensitive to this drift, as spike packets emitted by neuron cores on one board may arrive relatively late/early at synapse cores on another, reducing time available for their processing, or causing their arrival within the wrong timestep.

A master-slave beacon alignment protocol is therefore employed to alleviate board-to-board clock drift. The monitor core on every chip (see `M' core in Fig.~\ref{fig:mapping_and_data_transfer}) participates in the synchronisation, and instructs all neighbouring cores on the same chip how to adjust their clocks. A single master core sends a beacon packet to each of the slave cores every $2\, \text{s}$, with payload detailing the time of sending. Slave cores use this information, together with the packet arrival time, and compare the inter-beacon interval to the period elapsed on their timer. This allows each slave core to calculate a correction factor in terms of clock cycles, which is subsequently written to chip-level shared memory (SRAM). Each application core then reads this parameter and updates its timer period accordingly. Correction factors are typically small (less than a single clock cycle), and hence for accuracy corrections can be accumulated over multiple timer periods before being applied. 
While this mechanism accounts for clock drift, it does not ensure timers are in phase across the machine. When beginning a simulation, a start signal is sent from host to the SpiNNaker machine root monitor core (monitor core on chip 0,0), which then propagates this signal to the rest of the system. As the signal is routed from chip to chip, a small transmission time is encountered by each router ($500 \, \text{ns}$) and board-to-board link ($900 \, \text{ns}$), meaning chips farthest from the machine origin will receive the start signal latest. Furthermore, spike packets transmitted during simulation will encounter a similar transit time across the machine, meaning a spike sent at the beginning of the timer period by a core at the far edge of a machine will arrive at a core close to the origin with a relative delay of double the transit time. While the transit time is relatively small on the SpiNNaker machines discussed here (as it is proportional to machine size), in the extreme case this feature can cause packets to arrive during the wrong simulation timestep, and hence lead to unpredictable results. Phase alignment is therefore achieved using the start signal together with the chip-specific delay, which is calculated during boot and is proportional to the distance from the chip(s) with the largest start-signal transit time. For example, the chip at the origin receives the start signal instantaneously on beginning a simulation, but delays execution until the start signal has propagated to chips on the farthest edge of the machine. Conversely, a chip on the farthest edge of the machine acknowledges its location, and starts immediately on receiving the signal, hence causing all cores to start in phase. 

Measurements from the 12-board SpiNNaker system employing these alignment mechanisms found that core timers were aligned across boards to within $5 \, \mu\text{s}$. This ensures correctness in real-time SNN simulations using the discussed API, together with a timer period of $t_p = 100 \, \mu\text{s}$ (a requirement resulting from real-time execution of models with a simulation timestep of $\Delta t = 0.1 \, \text{ms}$).

\subsection{Energy \& Performance Profiling} \label{sec:energy_measurements}
A key target of neuromorphic computing is emulation of the low-power computational abilities of the brain. It is therefore useful to measure energy consumed during simulation execution, to contrast with both the brain and other computing platforms such as HPC and GPUs. Target measures characterising performance are energy to solution (energy consumed during the state propagation phase of simulation), and energy per synaptic event (where a synaptic event is defined as a single spike arriving at the synapse of an individual postsynaptic neuron). Recording such measures is a challenge due to the low power consumption of neuromorphic hardware, along with achieving consistency between measurements of different hardware platforms. While power monitoring equipment has been included on dedicated SpiNNaker boards for measurements from individual cores/chips, there are currently no instrumented large multi-board systems. In this work, a technique used in previous work \cite{Albada2018} is replicated, whereby an energy meter is used to record wall-socket power of the entire SpiNNaker system: SpiNNaker boards, communications switch, power supplies and cooling fans. The meter provides a reading accurate to $0.01 \, \text{kWh}$. A software controlled camera is instructed to take readings at the beginning and end of simulation, the difference between readings then defines the total energy consumed. This energy to solution is then converted into energy per synaptic event by dividing by the total model synaptic events, and assuming constant power consumption throughout the simulation period. 

All simulations and measurements reported here were executed on a 24-board SpiNNaker system, with machines allocated depending on simulation requirements. The cortical microcircuit model requires 6.6 boards for the DC input version, and 8.4 boards for the Poisson version, both of which result in allocation of a 12-board SpiNNaker machine. During simulation, unused chips remain in a booted but idle state, while unallocated boards (not part of the 12-board machine) remain powered off. In order to understand background power consumption relative to that of the SNN simulation, measurements are taken across three machine states: powered off; booted but idle; and actively running a simulation. In all cases readings are measured over a $12 \, \text{h}$ period in order to minimise error introduced through the energy meter resolution.


\section{Results}
The cortical microcircuit model described in Sec.~\ref{sec:background_and_motivation} was simulated on the SpiNNaker platform using the approach described in Sec.~\ref{sec:methods}. The following results document model and system performance in terms of real-time execution, accuracy and energy consumption.

\subsection{Real-Time Execution}
\begin{figure}
    \centering
    \includegraphics[width=0.75\textwidth]{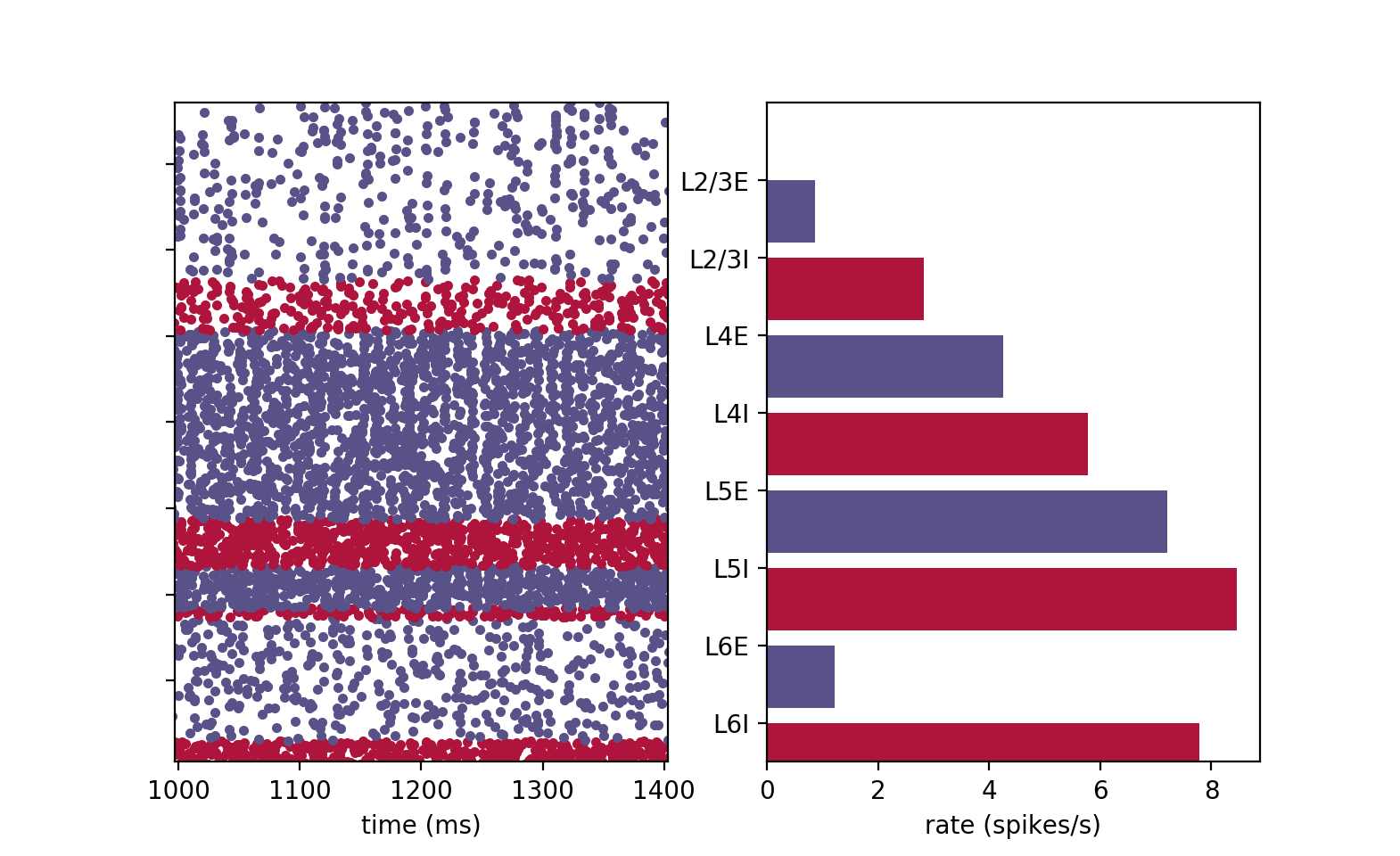}
    \caption{Results from \emph{true} real-time simulation of cortical microcircuit model with Poisson input on SpiNNaker. Left, $0.4 \, \text{s}$ of output spikes ($5 \%$ of total spikes plotted for clarity); and right, layer-wise mean population firing rates.}
    \label{fig:realtime_poisson_output}
\end{figure}

The cortical microcircuit model was executed for $10\,\text{s}$ of biological simulation time, with each $0.1 \, \text{ms}$ timestep evaluated in $0.1 \, \text{ms}$ CPU time, resulting in a wall-clock simulation time of $10\,\text{s}$, and \emph{true} real-time execution. Results for the Poisson version of this simulation are shown in Fig.~\ref{fig:realtime_poisson_output}.  Model initial conditions are relaxed from previous implementations \cite{Albada2018} to avoid excessive firing in the first simulation timestep due to neurons being initialised above threshold (see Fig.~\ref{fig:model_assessment}). 
Neuron membrane potentials are instead initialised to give population firing dynamics over the first simulation timesteps similar to steady-state activity. Figure~\ref{fig:realtime_poisson_output}, shows that these initial conditions do not affect the steady-state behaviour of the network, with good agreement observed between Figs.~\ref{fig:realtime_poisson_output}~\&~\ref{fig:microcircuit}. 

\begin{figure}[t!]
    \centering
    \includegraphics[width=\textwidth]{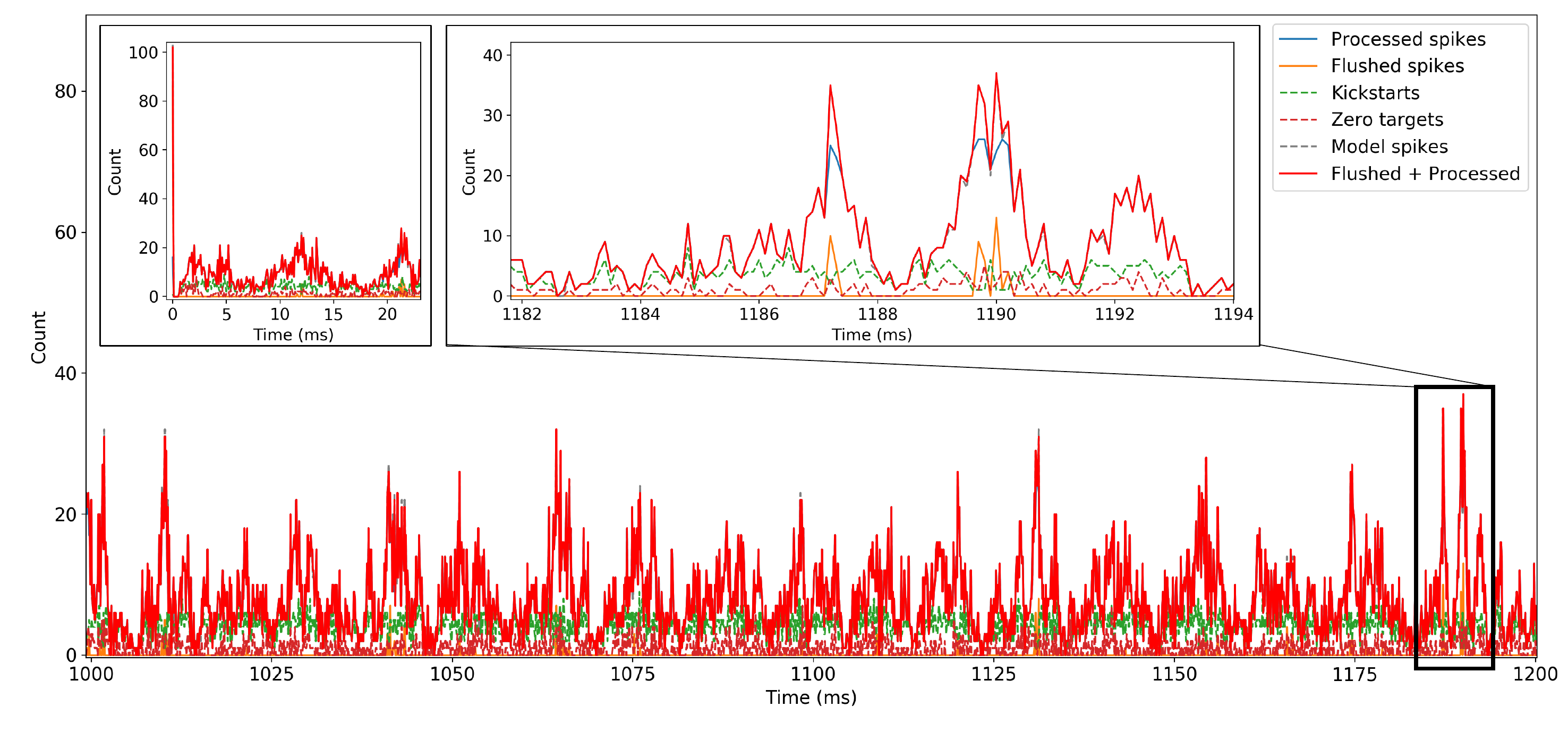}
    \caption{Synapse core profiling during simulation of the Poisson input cortical microcircuit. Total spike packets received per timestep plotted, together with total processed and flushed spikes: left inset details response around initial transient; while right inset details handling of extreme peak spike rates during steady-state oscillations.}
    \label{fig:realtime_SNN_results}
\end{figure}

To understand how the simulator coped with real-time execution, it is useful to examine the synapse cores as these are the only part of the ensemble to experience variable load during simulation. Profiling output from an upper excitatory synapse core from the L23E population is shown in  Fig.~\ref{fig:realtime_SNN_results} (core active in simulation results presented in Figs~\ref{fig:realtime_poisson_output}~\&~\ref{fig:accuracy_results}). A window of activity $1.0 \, \text{s} \leq t \leq 1.2 \, \text{s}$ is displayed, detailing the breakdown of incoming spikes to the core. The total spikes produced by the model which target this core is plotted by the dashed grey line, however this is difficult to view as it is obscured by the solid red line which details the total spikes handled by the core. This demonstrates correct operation of the neural ensemble throughout the simulation, as all spikes are accounted for in the appropriate timestep. 
The solid red line is made up of the total flushed spikes (orange curve) and processed spikes (blue curve). For the majority of the simulation no spikes are flushed, indicating good load balancing between the two excitatory synapse cores. However, the inset in Fig.~\ref{fig:realtime_SNN_results}, shows that during certain model oscillations incoming spike rates exceed the processing capacity of the core, and a number of spikes are flushed -- e.g. around $t=1190 \, \text{ms}$. Here the blue and red curves deviate from one another, demonstrating the peak processing capacity of the core to be approximately 26 spikes per timestep. Much larger flushes are observed during the initial timestep of the model (Fig.~\ref{fig:realtime_SNN_results}, right inset), as despite the improved initial conditions over $100$ spikes are delivered to upper excitatory synapse cores in the first timestep. The core is able to handle this extreme load, and maintains \emph{true} real-time execution by flushing excess spikes and proceeding to the following timestep. Also shown in Fig.~\ref{fig:realtime_SNN_results}, are curves detailing the number of spike packets targeting zero neurons on this particular synapse core (dashed crimson curve), and the number of times the spike processing pipeline was kick-started within a particular timestep (dashed green curve). The pipeline is often kick-started multiple times within a timer period, indicating that spike packets are received throughout the $t_{s,proc}$ window. It is also noted that the time penalty of pipeline kick-starts rarely impacts peak spike processing capacity, as when large numbers of spikes arrive within a timer period they automatically keep the pipeline active. However, spikes which target no neurons on the postsynaptic core do impact performance, particularly as the processing pipeline has already spent significant effort to discover this. While for L23E synapse cores this is not detrimental to performance, for postsynaptic populations receiving connections based on low probabilities, spikes not targetting any neurons can make up approximately half the spike traffic arriving at synapse cores. It is noted that this issue could be exacerbated in larger models with sparse connectivity, such as those containing spares long-range connections. However, there are a number of mechanisms which could be employed to remove this wasted effort, including increasing the neuron density per chip, and using hardware/software routing strategies to stop zero-target spike packets arriving at a synapse core, and/or minimising the effort required in their processing. 

\subsection{Simulation Accuracy}
While the cortical microcircuit model produced correct results when executed in real time, the flushing of unprocessed spikes on saturated synapse cores means information is lost. Table~\ref{tab:energy_measurements}, details the total synaptic events processed during real-time SpiNNaker simulations, with $9.135 \times 10^9$ and $9.343 \times 10^9$ reported for DC and Poisson input versions of the cortical microcircuit respectively. This compares with $9.416 \times10^9$ synaptic events in the NEST simulation with Poisson input, representing synaptic event losses within DC and Poisson SpiNNaker simulations of $2.98\%$ and $0.76\%$ respectively. It is thought that increased flushes occur during simulation of the DC version of the model, due mapping causing increased neuron density per chip in this configuration. This results in additional synapse cores on each chip, which in turn increases the chance of contention during synaptic row fetches (DMA~\textbf{A}), extending transfer times and reducing spike processing capacity. While this effect is relatively minor, it highlights the holistic nature of distributed systems design and the need for further research to optimise make-up of neural processing ensembles. 

\begin{figure}[t!]
    \begin{center}
    \includegraphics[width=\textwidth]{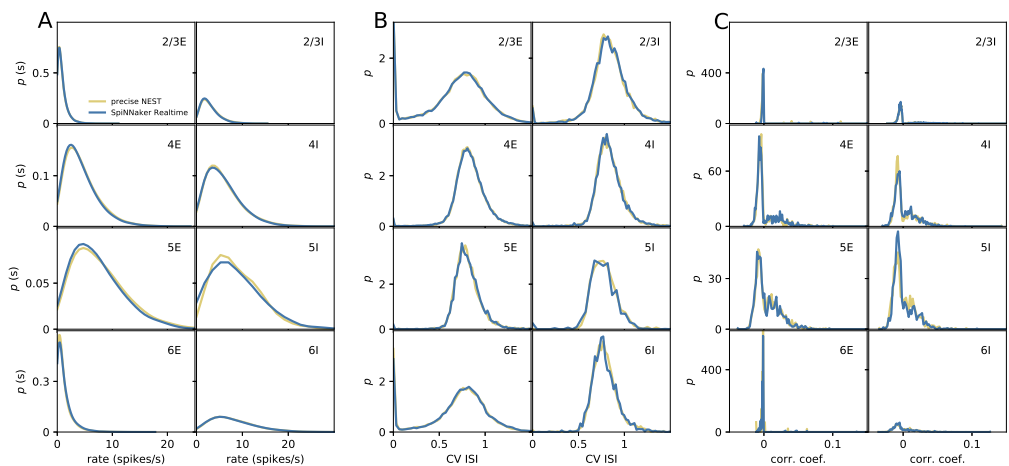}
    \caption{Comparison of spiking output from last $10\, \text{s}$ of cortical microcircuit simulations with Poisson input, executed using SpiNNaker in real time, and NEST at $3\times$ slow-down (all averaged over last $9\, \text{s}$ of simulation): A, single neuron firing rates; B, coefficient of variation of inter-spike interval; C, correlation coefficient between binned spiketrains.}
    \label{fig:accuracy_results}
    \end{center}
\end{figure}

To understand the effect of lost synaptic events, statistical analysis is performed on output spiketrains to compare quantitatively results with baseline NEST simulations. Resulting spiketrains are postprocessed to extract the within-population: distribution of firing rates; coefficients of variation of inter-spike interval; and correlation coefficients for binned spike trains. Where necessary, spiketrains are binned according to the Freedman-Diaconis rule. Results are displayed in Fig.~\ref{fig:accuracy_results}, for the real-time Poisson input SpiNNaker simulation (blue), together with corresponding results from the NEST simulator running at $3\times$ slow-down (yellow).
Good correlation between SpiNNaker and NEST results is observed across all metrics. The most significant variation is between the distributions of firing rates in the inhibitory layer 5 population, however this is in common with observations in previous work when changing simulator random number generator seeds \cite{Albada2018}. The results therefore confirm the model is executing correctly, and simulation accuracy has been preserved within the real-time heterogeneous parallelisation. 

\subsection{SpiNNaker Machine Mapping \& Energy}
\begin{figure}[t!]
    \centering
    \subfigure[]{\includegraphics[width=0.49\textwidth]{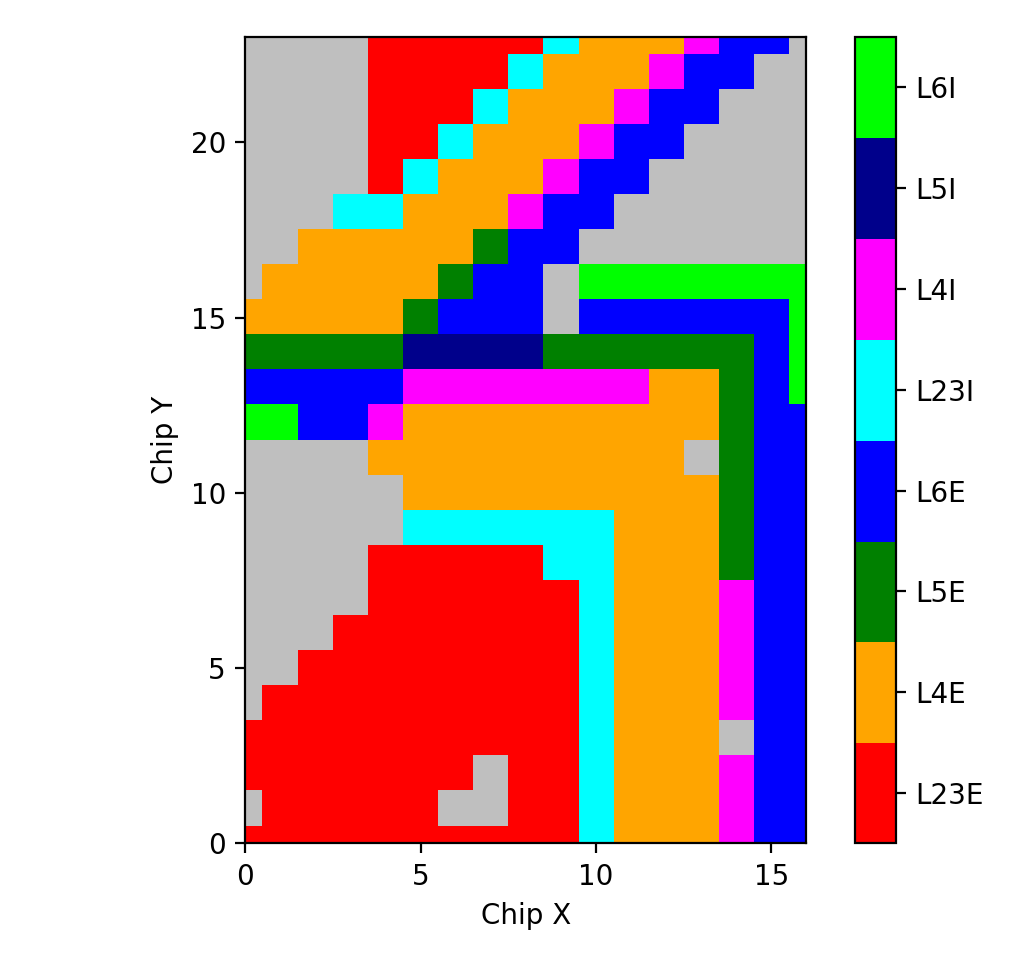}}
    \subfigure[]{\includegraphics[width=0.49\textwidth]{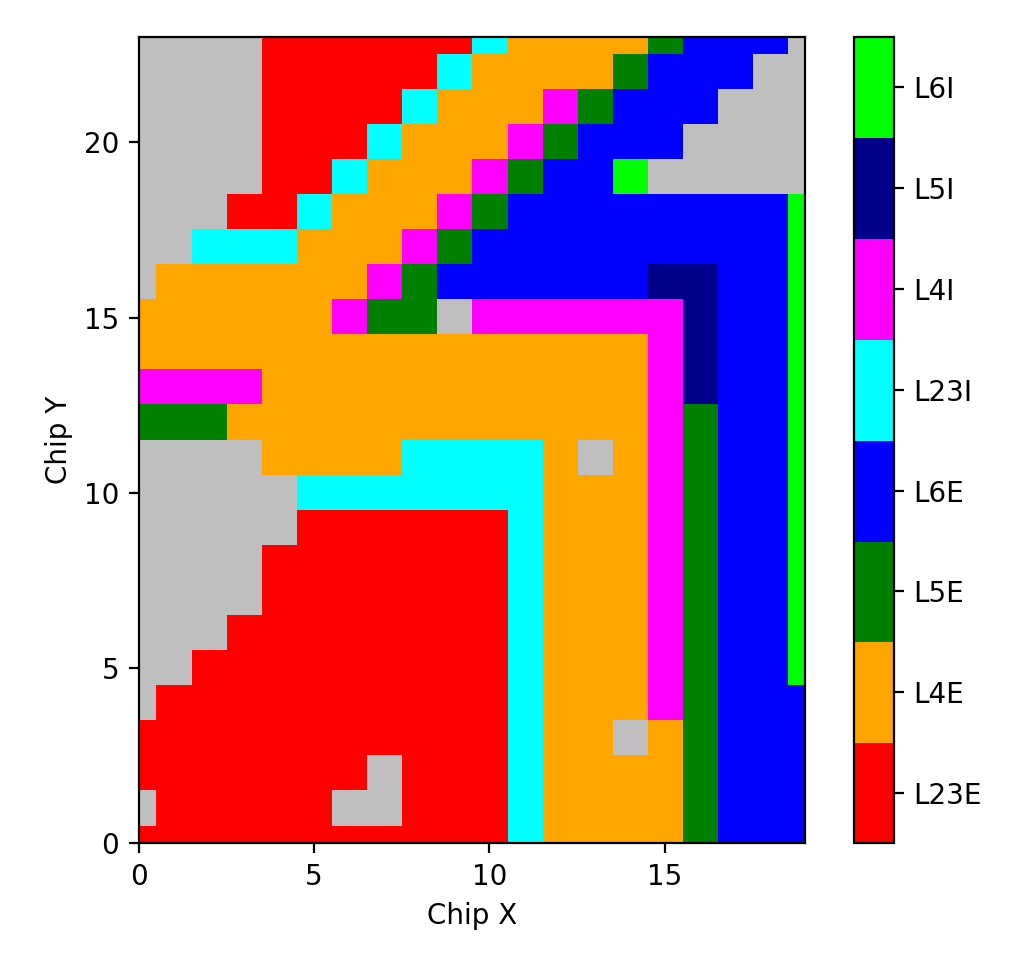}}
    \caption{Mapping of cortical microcircuit layer-wise excitatory and inhibitory populations to chip coordinates of the SpiNNaker machine: (a) DC input version, (b) Poisson input version.}
    \label{fig:population_machine_mapping}
\end{figure}

The model was executed on a 12-board SpiNNaker machine, with $576$ chips (with dimensions: 24 chips wide by 24 chips high) and $10085$ cores (total reduced from theoretical maximum due to failed cores present on active chips). This size of machine is provided by the SpiNNaker allocation software, which assigns square machines with dimensions incrementing by 3 boards with growing machine size. Mapping of the cortical SNN model populations to the SpiNNaker machine is detailed in Fig.~\ref{fig:population_machine_mapping}, where $(x,y)$ chip coordinates are coloured according to the assigned population. This shows that while a 12-board machine  is allocated by preprocessing software, the DC version of the model is mapped to 6.6 boards (a total of 4840 cores over 318 chips -- Fig.~\ref{fig:population_machine_mapping}(a)), while the Poisson version requires 8.4 boards (a total of 6050 cores over 404 chips -- Fig.~\ref{fig:population_machine_mapping}(b)). This is an increase in cores of $25\%$, due to the addition of a Poisson core to the neural processing ensemble. Note that for DC simulations, four ensembles can be mapped to a single chip, giving a neuron density of 256 neurons per chip. The Poisson input version of the model maps three ensembles to a chip, leading to a neuron density of 192 neurons per chip. 

The 12-board machine is allocated from a self-contained 24-board system, which is configured with wrap-around connectivity at its top and bottom edges, reducing the shortest path length for spike packets traversing the machine. This machine continuity at top and bottom edges is visible in Fig.~\ref{fig:population_machine_mapping}, where a radial placer is used to map partitioned model populations to SpiNNaker chips, resulting in the outward spiralling of populations from the machine origin (chip $(0,0)$). This arrangement works favourably in terms of spike packet distribution to all synapse processing cores, due to the interleaving of excitatory and inhibitory populations. This ensures that spike packets from adjacent neuron cores arrive quickly, while packets from distant neuron cores will arrive later, thus helping kick-start the spike processing pipeline early in the timer period, and helping it remain active as further packets arrive. 

Energy use was measured according to the methods described in Sec.~\ref{sec:methods}\ref{sec:energy_measurements}, and is reported for a range of configurations in Tab.~\ref{tab:energy_measurements}. The first three configurations report the energy required to power the system for 12 hours, enabling precision in power recording equipment through accumulation of readings over long time windows. The `System Only' configuration details the base power of the machine with all SpiNNaker boards powered off, and therefore represents auxiliary system components including cooling fans, communication switch, and rack power supplies. The `12 boards booted' configuration captures the energy used by the 12-board SpiNNaker machine in its powered on and booted state, demonstrating an increase of $5.3\times$ relative to the powered-off machine. 
\begin{table}[]
    \centering
    \begin{tabular}{m{3.5cm} m{1.75cm} m{2cm} m{1.75cm} m{2.65cm}} \hline
        Configuration & Wall-Clock Time &  Total Energy (kWh) & Synaptic Events & Energy per synaptic event ($\mu\text{J/syn-event}$)  \\ \hline
        System Only & $12 \,\text{h}$ & 0.93 & -- & --  \\
        12 booted boards & $12 \,\text{h}$ & 4.93 & -- & --  \\
        Cortical microcircuit, DC input, real-time & $12 \,\text{h}$ &  6.59 & -- & --\\
        Cortical microcircuit, Poisson input, real-time & $12 \,\text{h}$ &  7.11 & -- & --\\
        Cortical microcircuit, DC input, real-time & $10 \,\text{s}$ &  $0.001525$* & $9.135\times10^9$ & $0.601$ \\ 
        Cortical microcircuit, Poisson input, real-time & $10 \,\text{s}$ &  $0.001646$* & $9.343\times10^9$ & $0.628$ \\ \hline
    \end{tabular}
    \caption{Energy measurements from 24-board SpiNNaker system used to execute cortical microcircuit simulations (*quantity estimated from extended duration simulation).}
    \label{tab:energy_measurements}
\end{table}
Results for both the DC and Poisson input versions of the cortical microcircuit simulation are then reported (also executed for $12\,\text{h}$). The DC version consumes $1.34\times$ more energy than the booted but idle SpiNNaker system, demonstrating that total energy consumption contributed by SNN simulation is approximately $25\%$. The Poisson version of the model consumes $1.44\times$ more energy than the booted but idle configuration, a slight increase relative to the DC version due to the additional chips/cores used, demonstrating that $30\%$ of total energy consumed is used by the SNN simulation with Poisson input.

The bottom two rows of Tab.~\ref{tab:energy_measurements}, show estimated energy consumption of $10 \, \text{s}$ cortical mircocircuit simulations with DC and Poisson input. This value is then used, together with the recorded synaptic events from each simulation, to estimate energy per synaptic event. This is calculated at $0.601 \, \mu\text{J}$ and $0.628 \, \mu\text{J}$ per synaptic event for the respective DC and Poisson simulations. This represents an order of magnitude reduction relative to previous results for both SpiNNaker and HPC ($5.9\, \mu\text{J}$ and $5.8\, \mu\text{J}$ per synaptic event respectively \cite{Albada2018}). It also demonstrates how the relatively low-tech SpiNNaker processors, based on a $130\, \text{nm}$ CMOS technology much older than that used in the comparator HPC or GPU systems, compete favourably with a range of modern GPUs running optimised SNN libraries ($0.3 - 2.0\, \mu\text{J/syn-event}$ \cite{Knight2018}).


\section{Conclusion}
This work presents the first real-time execution of a published large-scale cortical microcircuit model. This result surpasses previously published results in terms of processing speed, with optimal performance relative to real time reported at $3\times$ slow-down for an HPC-based simulator \cite{Albada2018}, and $2\times$ slow-down for GPUs running optimised SNN libraries \cite{Knight2018}. Real-time processing was achieved through the use of SpiNNaker neuromorphic hardware, together with a heterogeneous programming model mapping neural processing to an ensemble of cores, each tackling different simulation operations. This approach enabled parallelisation of incoming spike traffic, and hence additional synapse-focused cores to be added to the ensemble to handle simulation requirements. Energy use for the proposed approach is measured at $0.628, \mu\text{J/syn-event}$, surpassing published HPC performance figures by an order of magnitude \cite{Albada2018}, and giving comparable consumption to modern GPU hardware \cite{Knight2018}. 

Neuron density per chip is reduced significantly from initial SpiNNaker design targets (256 neurons per chip for DC version, and 192 for Poisson version of the model), however the model specification is also increased in complexity from initial design targets, containing an order of magnitude more synaptic connections per neuron, and requiring simulation timesteps of $\Delta t = 0.1 \, \text{ms}$ -- also an order of magnitude less than original predictions. Energy consumption figures therefore  stand to be improved dramatically by further research and software optimisations, which could allow simulation of increased numbers of neurons per chip, and mapping of the cortical microcircuit to a much smaller SpiNNaker machine.  Likely candidates for optimisation include speeding up of neuron state updates, and combining of Poisson input source code into neuron core applications, freeing up multiple cores per chip to run additional neural processing ensembles. However, it should highlighted that SpiNNaker simulations are already operating across multiple boards, and hence have overcome the communication bottleneck limiting further performance gains from HPC and GPUs. The parallelisation of spike communication presented here therefore not only improves speed and energy efficiency at this scale of model, but stands to scale further with model size and the use of additional SpiNNaker hardware, preserving both real-time processing and energy per synaptic event levels.  

System correctness and robustness is also demonstrated, with all spikes accounted for within the spike processing pipeline, and multiple $12\, \text{h}$ simulations performed successfully. This deterministic and predictable nature, coupled with extended duration simulations executed in real time, opens the door for a new paradigm in neuroscience research. Neuromorphic hardware now offers the potential to study long-term effects in spiking neural network models representing brain activity, both at the same speed as they occur in biology, and at scales not measurable through physical experiments. It also opens the door to deploying large-scale SNNs in real-time neurorobotics applications, and exploring applications of large-scale brain-inspired artificial intelligence.

\enlargethispage{20pt}

\dataccess{The data and code used to generate the results presented in this paper are available from the SpiNNaker software stack: \href{https://github.com/SpiNNakerManchester}{https://github.com/SpiNNakerManchester}, using branches \emph{realtime\_cortical\_microcircuit}.}

\aucontribute{LP co-designed and built the SpiNNaker software implementation of the heterogeneous programming model. ADR implemented the improved synaptic matrix generation and access, along with the board-to-board drift alignment protocol. AG implemented the Poisson application delivering neuron input through shared memory. LAP implemented the board-to-board phase alignment, and assisted with low-level SpiNNaker application code and hardware interfaces. CB implemented the updated packet key structure and key generation mechanisms. SBF leads the SpiNNaker group and supervised the research. OR conceived the study, performed the analysis of the cortical microcircuit model, co-designed the SpiNNaker software implementation of the heterogeneous programming model, ran the simulations and drafted the manuscript. All authors read, commented and approved the final manuscript.}

\competing{SBF is a founder, director and shareholder of Cogniscience Ltd, which owns SpiNNaker IP. LAP and AGDR are shareholders of Cogniscience Ltd.}

\funding{The design and construction of the SpiNNaker machine was supported by EPSRC (the UK Engineering and Physical Sciences Research Council) under grants EP/D07908X/1 and EP/G015740/1, in collaboration with the universities of Southampton, Cambridge and Sheffield and with industry partners ARM Ltd, Silistix Ltd and Thales. Ongoing development of the software, including the work reported here, is supported by the EU ICT Flagship Human Brain Project (H2020 785907), in collaboration with many university and industry partners across the EU and beyond. LP is funded by an EPSRC DTA studentship in the Department of Computer Science.}

\ack{The authors would like to thank Hanjia Jiang, Sacha van Albada and Marcus Diesman of FZ J{\"u}lich, Germany, for providing NEST simulation results of the cortical microcircuit model, and with assistance in generating the updated initial conditions used in neuromorphic simulations reported here. We would also like to thank Mantas Mikaitis and Robert James of the Advanced Processor Technologies group at the University of Manchester, for assistance with optimisation of SpiNNaker application code, and discussions on synaptic parallelisation and improving performance in large-scale SNNs.}



\end{document}